\documentclass[conference]{IEEEtran}
\IEEEoverridecommandlockouts
\usepackage{cite}
\usepackage{amsmath,amssymb,amsfonts}
\usepackage{mathtools}
\usepackage{graphicx}
\DeclareGraphicsExtensions{.eps,.jpeg,.png,.pdf}
\usepackage{textcomp}
\usepackage{xcolor}

\usepackage{hyperref}
\usepackage{siunitx}
\usepackage{stfloats}
\usepackage{multirow}

\usepackage{amsthm}
\usepackage{pdfpages}
\usepackage{layout}

\def\BibTeX{{\rm B\kern-.05em{\sc i\kern-.025em b}\kern-.08em
    T\kern-.1667em\lower.7ex\hbox{E}\kern-.125emX}}

\newcommand{\udq}{\boldsymbol{u}_\mathrm{dq}}
\newcommand{\idq}{\boldsymbol{i}_\mathrm{dq}}

\newcommand{\Ldq}{\boldsymbol{L}_\mathrm{dq}}
\newcommand{\Psip}{\psi_{\mathrm{p}}}
\newcommand{\Psidq}{\boldsymbol{\psi}_{\mathrm{dq}}}
\newcommand{\Psid}{\psi_{\mathrm{d}}}
\newcommand{\Psiq}{\psi_{\mathrm{q}}}
\newcommand{\J}{\boldsymbol{J}}
\newcommand{\UDC}{U_{\mathrm{DC}}}
\newcommand{\wel}{\omega_\mathrm{el}}
\newcommand{\eel}{\varepsilon_\mathrm{el}}
\newcommand{\ud}{u_\mathrm{d}}
\newcommand{\uq}{u_\mathrm{q}}
\newcommand{\id}{i_\mathrm{d}}

\newcommand{\iq}{i_\mathrm{q}}
\newcommand{\Ld}{L_\mathrm{d}}
\newcommand{\Lq}{L_\mathrm{q}}
\newcommand{\Ts}{\mathrm{T_s}}
\newcommand{\Rs}{R_\mathrm{s}}
\newcommand{\h}{\frac{1}{2}}
\newcommand{\s}{\frac{\sqrt{3}}{2}}
\newcommand{\sthree}{\frac{1}{\sqrt{3}}}
\newcommand{\vn}{\boldsymbol{v}_n}
\newcommand{\san}{s_{\mathrm{a},n}}
\newcommand{\sbn}{s_{\mathrm{b},n}}
\newcommand{\scn}{s_{\mathrm{c},n}}
\newcommand{\idk}{i_{\mathrm{d},k}}
\newcommand{\iqk}{i_{\mathrm{q},k}}
\newcommand{\eelk}{\varepsilon_{\mathrm{el},k}}
\newcommand{\sink}{\sin(\varepsilon_{\mathrm{el},k})}
\newcommand{\cosk}{\cos(\varepsilon_{\mathrm{el},k})}
\newcommand{\idkk}{i_{\mathrm{d},k+1}}
\newcommand{\iqkk}{i_{\mathrm{q},k+1}}

\newcommand{\FESystemMatrix}{\begin{bmatrix} -\frac{\Rs}{\Ld} & \frac{\Lq}{\Ld}\wel & \frac{\UDC}{2\Ld}\left(\sthree \sbn - \sthree \scn \right) & \frac{\UDC}{2\Ld}\left(\frac{2}{3}\san-\frac{1}{3}\sbn-\frac{1}{3}\scn \right) & 0 \\ -\frac{\Ld}{\Lq}\wel & -\frac{\Rs}{\Lq} & \frac{\UDC}{2\Lq}\left(\frac{2}{3}\san-\frac{1}{3}\sbn-\frac{1}{3}\scn \right) & \frac{\UDC}{2\Lq}\left(\sthree \sbn - \sthree \scn \right) &  -\frac{\Psip}{\Lq}\wel \\ 0&0&0&\wel&0 \\ 0&0&-\wel&0&0 \\ 0&0&0&0&0 \end{bmatrix}}
\begin{document}

\title{Data Set Description: Identifying the Physics Behind an Electric Motor -- Data-Driven Learning of the Electrical Behavior (Part II)
\thanks{This work was funded by the German Research Foundation (DFG) under the reference number BO 2535/11-1.}
}

\author{

\IEEEauthorblockN{S{\"o}ren Hanke\IEEEauthorrefmark{1},
									Oliver Wallscheid, 
									Joachim B{\"o}cker}
													
\IEEEauthorblockA{Department of Power Electronics and Electrical Drives, Paderborn University, 33095 Paderborn, Germany,\\\IEEEauthorrefmark{1}E-mail: hanke@lea.upb.de}
% \IEEEauthorblockA{\IEEEauthorrefmark{2}Applied Mathematics, Paderborn University, 33095 Paderborn, Germany,\\ speitz@math.upb.de, dellnitz@upb.de}
}

\maketitle

\begin{abstract}
A data set was recorded to evaluate different methods for extracting mathematical models for a three-phase permanent magnet synchronous motor (PMSM) and a two-level IGBT inverter from measurement data.
It consists of approximately 40 million multidimensional samples from a defined operating range of the drive.
This document describes how to use the published data set \cite{Dataset} and how to extract models using introductory examples.
The examples are based on known ordinary differential equations, the least squares method or on (deep) machine learning methods.
The extracted models are used for the prediction of system states in a model predictive control (MPC) environment of the drive.
In case of model deviations, the performance utilizing MPC remains below its potential.
This is the case for state-of-the-art white-box models that are based only on nominal drive parameters and are valid in only limited operation regions. 
Moreover, many parasitic effects (e.g. from the feeding inverter) are normally not covered in white-box models.
In order to achieve a high control performance, it is necessary to use models that cover the motor behavior in all operating points sufficiently well.
\end{abstract}

\section{Preliminary remark}
The description of the data set consists of two parts. Part I gives a simplified introduction to the system behind the data and explains how to use the data set (\url{https://arxiv.org/abs/2003.07273}) \cite{Description_Part_I}.

Part II (this document) explains the system in more details, covers some basic approaches on how to extract models and discusses also a possible way to get a balanced data set where the samples are evenly distributed in a subset used for (deep) machine learning (ML) methods (\url{https://arxiv.org/abs/2003.06268}) \cite{Description_Part_II}.

%Without any further reference, parts of the following explanations are based on or directly taken from the author's publications \cite{Hanke_19_2} and \cite{Hanke_20_1}. 

\section{Introduction}

The idea of a data-driven modeling approach is to extract governing equations from measurement data.
These equations can also incorporate effects that can hardly be considered in a white-box modeling approach relying only on domain-specific expert knowledge.
In a permanent magnet synchronous motor (PMSM), which is often used as a traction motor in electric vehicles, such effects can be the strong magnetic saturation of the inductances, common and differential mode capacitive influences or temperature dependencies.

% Models can be extracted from the data online, during runtime of the drive system, or offline from recorded measurement data.

In the remainder of this paper, the drive system and its basic white-box modeling will be explained first (Sec. \ref{sec:Drive system}).
In Sec. \ref{sec:FCS-MPC}, the usage of the models within the model predictive control (MPC) is explained.
Basic approaches for the extraction of models from data, including introductory examples, are discussed in Sec. \ref{sec:Prediction_models}.
With this basic understanding of the drive system, the recording and the data driven model extraction the characteristics of the data set are explained and analyzed (Sec. \ref{sec:data set}).

\section{Drive system}
\label{sec:Drive system}

The structure of the drive system including the FCS-MPC with the used prediction models is shown in Fig. \ref{fig:drive_system}.
The ordinary differential equations (ODE) of the PMSM in the rotor-flux oriented dq-system are given by (first principle model):
	\begin{figure}[ht]
		\centering
		\includegraphics[]{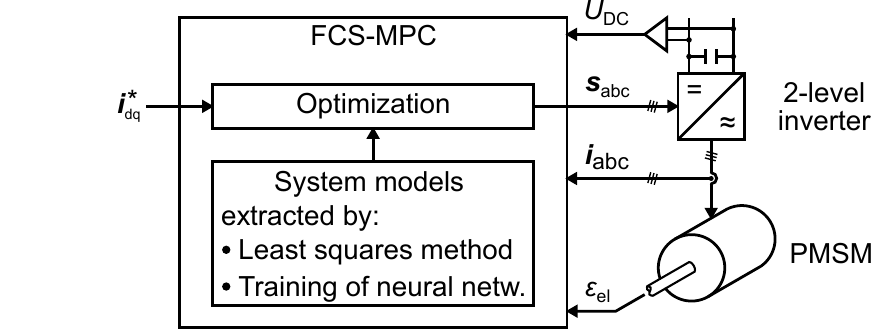}
		\caption{Structure of the drive system}
		\label{fig:drive_system}
	\end{figure}

				\begin{figure*}[!b]
				
				\hrulefill
				\vspace*{0pt}
				
				\normalsize
				\setcounter{equation}{5}
				
				\begin{equation}
				\begin{split}
				\underbrace{\vphantom{\FESystemMatrix} \frac{\mathrm{d}}{\mathrm{d}t} \begin{bmatrix} i_{\mathrm{d}} \\ i_{\mathrm{q}} \\ \sin(\eel) \\ \cos(\eel) \\ 1 \end{bmatrix}}_{\textstyle \frac{\mathrm{d}}{\mathrm{d}t}{\boldsymbol{x}}} &= \underbrace{\FESystemMatrix}_{\textstyle  \boldsymbol{A}_n} \underbrace{\vphantom{\FESystemMatrix} \begin{bmatrix} \id \\ \iq \\ \sin(\eel) \\ \cos(\eel) \\ 1 \end{bmatrix}}_{\textstyle \boldsymbol{x}}
				\label{eq:ODE_PMSM_autonomous}
				\end{split}
				\end{equation}
				
				\setcounter{equation}{0}
				
				\end{figure*}		
	
	\begin{equation}
	\begin{split}
			\udq   &= \Rs \idq + \wel \J \Psidq + \frac{\mathrm{d}}{\mathrm{dt}} \Psidq, \\
			\J     &= \begin{bmatrix} 0 & -1 \\ 1 & 0 \end{bmatrix},\\
			\Psidq &= \begin{bmatrix} \Psid(\id, \iq, \eel) \\ \Psiq(\id, \iq, \eel) \end{bmatrix}.\\
	\label{eq:ODE_PMSM_general}			
	\end{split}
	\end{equation}	
	
Above, $\udq=\begin{bsmallmatrix} \ud & \uq \end{bsmallmatrix}^\top$ is the stator voltage, $\Rs$ is the stator resistance, $\idq$ is the stator current, $\wel$ is the electrical angular frequency, $\Psidq$ is the flux linkage, and $\eel$ is the electrical rotation angle of the PMSM. The flux linkage depends on the currents and the rotation angle to account for magnetic saturation effects and effects like cogging torque due to the mechanical construction of the machine's rotor and stator.

The the dq-coordinate system is a typically used coordinate system for a mathematical modeling in the motor control domain. 
Modeled in this system, the PMSM is very similar to the classical DC motor where the control is rather simple to realize.
Thus proven control concepts of the DC motor could easily be used as a basis for a PMSM controller.
However, the dq-system variables resulting from the state transformation of the physical system variables cannot be measured directly in the system.

%However, after the state transformation of the physical system variables into the dq-system, there is no direct physical interpretation of the dq-system variables.
%However, after the state transformation of the physical system variables into the dq-system, there is no direct measurement of the dq-system variables.
%With consideration of the following assumptions and simplifications \eqref{eq:ODE_PMSM_assumptions}, a basic PMSM model can be derived \eqref{eq:ODE_PMSM_basic}:

Assuming
	\begin{equation}
	\begin{split}
			\Psidq &= \begin{bmatrix} \Psid \\ \Psiq \end{bmatrix} = \begin{bmatrix} \Ld \id + \Psip \\ \Lq \iq \end{bmatrix} = \Ldq \idq + \begin{bmatrix} \Psip \\ 0 \end{bmatrix}, \\
			\Ldq   &= \begin{bmatrix} \Ld & 0 \\ 0 & \Lq\end{bmatrix},\\
	\label{eq:ODE_PMSM_assumptions}			
	\end{split}
	\end{equation} a basic PMSM model can be derived:		
	\renewcommand{\arraystretch}{1.3}
	\begin{equation}
	\frac{\mathrm{d}}{\mathrm{d}t}\boldsymbol{i}_{\mathrm{dq}} = \begin{bmatrix} -\frac{\Rs}{\Ld} & \frac{\Lq\wel}{\Ld}\\	-\frac{\Ld\wel}{\Lq} & -\frac{\Rs}{\Lq} \end{bmatrix} \idq + \Ldq^{-1}\udq+ \begin{bmatrix} 0\\ -\frac{\Psip\wel}{\Lq}\end{bmatrix}.
  \label{eq:ODE_PMSM_basic}
	\end{equation}
	\renewcommand{\arraystretch}{1.0}
		
Above, $\Ldq$ is the inductance matrix and $\Psip$ is the permanent magnet flux linkage. 
In this basic first principle ODE model, saturation effects and angle dependencies of the flux are neglected.
Moreover, parasitic effects such as inductive and capacitive influences of the cabling between inverter and motor or motor-specific construction asymmetries are not covered since those cannot be easily introduced to the white-box model.

The voltage $\udq$ is supplied to the motor by the 2-level voltage source inverter and can be mathematically expressed by
	\begin{equation}
	\begin{split}
		&\udq = \boldsymbol{Q}(\eel)\frac{\UDC}{3}\begin{bmatrix} 1 & -\h & -\h \\ 0 & \s & -\s \end{bmatrix} \vn,\\
		&\boldsymbol{Q}(\eel) = \begin{bmatrix} \cos(\eel) & \sin(\eel) \\ -\sin(\eel) & \cos(\eel) \ \end{bmatrix}.
	\end{split}
	\end{equation}

Here, $\boldsymbol{Q}$ is the rotation matrix, $\UDC$ is the DC-link voltage and $\vn$ the vector comprising the switching state of each phase for the eight elementary vectors of the 2-level inverter:
	\begin{equation}
	\begin{aligned}
	\vn &=\begin{bmatrix} \san & \sbn & \scn \end{bmatrix}^\mathrm{T},\ n=1,\ldots,8\\
	    &\ \mathrm{with}\ \san,\ \sbn,\ \scn\in\left\{+1;-1\right\}.
	\label{eq:vn}
	\end{aligned}
	\end{equation}		
		
Each elementary vector $\vn$ defines an autonomous system with its three switching states $\san$, $\sbn$, and $\scn$ (Tab. \ref{tab:elementary_vectors}).
The index $n$ denotes the corresponding autonomous system.
\begin{table}[!ht]
\centering
\caption{Inverter switching states of the eight autonomous systems}
\label{tab:elementary_vectors}
\begin{tabular}{c|ccc|}
\multirow{2}{*}{$n$} & \multicolumn{3}{c|}{$\vn$} \\ \cline{2-4}
                     & $\san$      &  $\sbn$     & $\scn$ \\ \hline
                  1  & -1          & -1          & -1     \\
                  2  & +1          & -1          & -1     \\
                  3  & +1          & +1          & -1     \\
                  4  & -1          & +1          & -1     \\
                  5  & -1          & +1          & +1     \\
                  6  & -1          & -1          & +1     \\
                  7  & +1          & -1          & +1     \\
									8  & +1          & +1          & +1     \\
\end{tabular}
\end{table}

To include the switching state information directly as part of the plant model, the basic model of the PMSM can also be expressed in the form given by (\ref{eq:ODE_PMSM_autonomous}) with the system matrix $\boldsymbol{A}_n$. Here, the state vector $\boldsymbol{x}$ includes the stator currents, the sine and cosine of the rotor angle and a constant value.
The constant value is required to include the induced voltage term of the q-current equation into the matrix as well.

Since the elementary vectors $\boldsymbol{v}_1$ and $\boldsymbol{v}_8$ lead to the same system matrix (both applying zero voltage to the motor), it is sufficient to consider only one of these two vectors in the following. Therefore, the data set presented in the following will only contain samples for the vectors $\boldsymbol{v}_1$ to $\boldsymbol{v}_7$.

\section{Finite-control-set model predictive control}
\label{sec:FCS-MPC}

To solve an optimal control problem on a receding prediction horizon is the basic concept of the model predictive control. 
In each controller cycle, a mathematical model of the plant in conjunction with a cost function is used to find an optimal sequence of the actuating variables that minimizes the costs. 
Applying the first element of the sequence to the plant and repeating the optimization on the basis of new measurements of the system states, closes the control loop. 
Generally, the control error is one of the objectives in the cost function. %\textcolor[rgb]{1,0,0}{direkt aus \cite{Hanke_19_2} übernommen}

Since the controller is implemented on a digital hardware, the controller computations must be performed in a discrete-time manner, which implies the need for discretization if the models to be used are based on ODEs.

\begin{figure}[ht]
		\centering
		\includegraphics[]{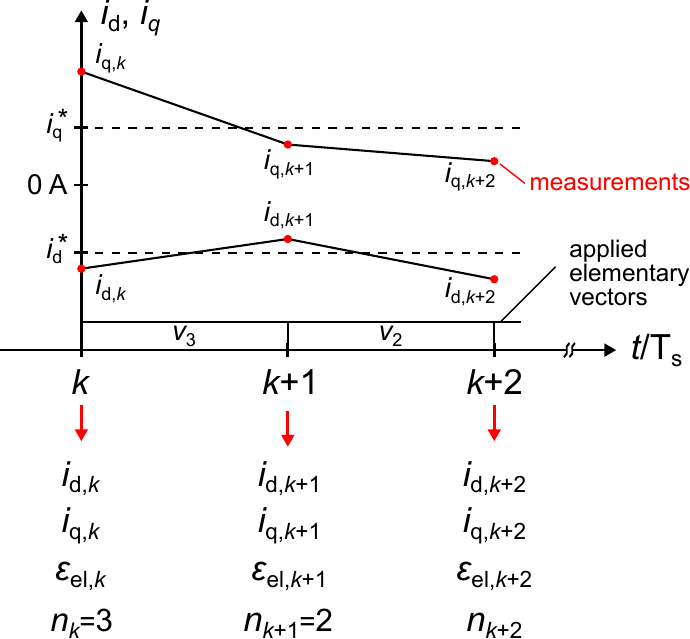}
		\caption{FCS-MPC: arbitrary curve shape with highlighted measurements}
		\label{fig:FCS_MPC_principle}
	\end{figure}

The finite-control-set model predictive control (FCS-MPC) selects the actuating variables among the eight possible autonomous systems $n$ which are defined by the elementary vectors of the inverter.
Fig. \ref{fig:FCS_MPC_principle} shows an arbitrary curve shape of the dq-currents when using an FCS-MPC and highlights the measurements carried out at each control cycle.

Since the FCS-MPC yields constant inputs within each controller cycle, a linear FCS-MPC predicts the future system states for the time point $(k+1)\Ts$ based on a discrete-time model.
In the context of (deep) machine learning (ML) methods (like Artificial Neural Networks (ANNs), Gradient-Boosting-Machines (GBM) or decision trees), usually nonlinear models (\ref{eq:prediction_ANN}) are used for an approximation of the plant behavior. 
The states are comprised in the vector $\boldsymbol{x}$. 

	\setcounter{equation}{6}
	
	\begin{equation}
		\hat{\boldsymbol{x}}_{k+1} = f_n\left(\boldsymbol{x}_{k}\right)
		\label{eq:prediction_ANN}
		\end{equation}	

With least squares (LS) methods, often linear discrete-time models (\ref{eq:prediction}) are used for the regression.
Here, the models for the drive's autonomous systems are denoted as $\boldsymbol{K}_n$.

	\begin{equation}
	\hat{\boldsymbol{x}}_{k+1} = \boldsymbol{K}_n\ \boldsymbol{x}_{k}
	\label{eq:prediction}
	\end{equation}

In comparison, nonlinear models provide a higher number of degrees of freedom and, thus, allow a more precise approximation of effects like saturation.
However, there is always a trade-off between prediction accuracy and computational complexity of a given model topology.
	
The main interest is in the prediction of the two currents $\id$ and $\iq$ which are part of the defined system states $\boldsymbol{x}$.
The increment in the rotation angle $\eel$ can be calculated easily, as the rather slow varying rotational speed $\wel$ is tracked by an phase locked loop (PLL) and the time increment is known to be $\Ts$.
The latter is assumed to be constant during all experiments i.e. the FCS-MPC is operated at a fixed controller cycling time.

\section{Prediction models for Finite-control-set model predictive control}
\label{sec:Prediction_models}

The plant models can be obtained by different approaches. Some of them which will be also described in the following are:

\begin{description}
	\item[a)] discretization of the white box ODE-based models (Sec. \ref{sec:Drive system}), 
	\item[b)] extraction from measurement data by using a least squares (LS) method,
	\item[c)] extraction from measurement data by (deep) machine learning (ML) methods.\\
\end{description}

An explanation on how to evaluate the accuracy of obtained models including an appropriate cost function is given in Part I of the data set description.\\

\noindent \textbf{Approach a) Discretization of ODE-based models}\\
The usage of the basic plant model for approach a) results in a discretization of the continuous-time ODE system \eqref{eq:ODE_PMSM_autonomous}.
The discrete-time form is obtained by using the transition matrix $\boldsymbol{\Phi}_n$ with a fixed time increment of $\Ts$:
	\begin{equation}
	\underbrace{\boldsymbol{x}((k+1)\Ts)}_{\boldsymbol{x}_{k+1}} = \boldsymbol{\Phi}_n(\Ts)\ \underbrace{\boldsymbol{x}(k \Ts)}_{\boldsymbol{x}_{k}}.
	\label{eq:discrete}
	\end{equation}
	
The transition matrix can be calculated with a series expansion, also known as matrix exponential:
	\begin{equation}
	\begin{split}
	\boldsymbol{\Phi}_n(\Delta t) = e^{\boldsymbol{A}_n \Delta t}\ &= \sum_{\nu=0}^\infty \frac{(\boldsymbol{A}_n \Delta t)^\nu}{{\nu}!} \\
	                                                &= \boldsymbol{I} + \boldsymbol{A}_n \Delta t + \frac{(\boldsymbol{A}_n \Delta t)^2}{2!} + \cdots
	\label{eq:transition_matrix}
	\end{split}
	\end{equation}	
																									
Usually, the linear discrete-time FCS-MPC models $\boldsymbol{K}_n$ built according to approach a) are an approximation of the transition matrix $\boldsymbol{\Phi}_n(\Ts)$ using the series expansion up to the 1st-order term:
	\begin{equation}
	\boldsymbol{K}_n       = \boldsymbol{I} + \boldsymbol{A}_n \Ts \approx \boldsymbol{\Phi}_n(\Ts).
	\label{eq:white-box-approximation}
	\end{equation}

However, this type of discretization assumes constant parameters in the white-box model $\boldsymbol{A}_n$, as they result from the simplification \eqref{eq:ODE_PMSM_assumptions}.
The more general model \eqref{eq:ODE_PMSM_general} can also be expressed as autonomous systems without the need for additional elements in $\boldsymbol{x}$.
But then the elements of the matrices $\boldsymbol{A}_n$ are dependent on the dq-currents and the rotor angle, resulting in a parameter-variant system.
The discretization itself then also depends on these parameters and the series expansion \eqref{eq:transition_matrix} or its approximation would therefore have to be recalculated for each controller cycle, which also results in an increased computational burden.\\

%The discretization itself then also depends on these parameters and cannot be determined by a series expansion as in \eqref{eq:transition_matrix}.\\

\noindent \textbf{Approach b) Least-squares-based models}\\
The extraction of the matrices $\boldsymbol{K}_n$ from data is one solution to account for the parameter-variant characteristic of the system.
Here, also effects which are not or only partially considered in a mathematical white-box model can be covered by the data-driven approaches b) and c).

Using the least squares (LS) approach, a multiple linear regression can be conducted.
During operation of the drive, the vector $\boldsymbol{x}$ is measured at $k\Ts$ and $(k+1)\Ts$ and the used elementary vector $\vn$ between these points in time is known.
%Thus, for each elementary vector $\vn$, several pairs, each consisting of $\boldsymbol{x}_{k,n}$ and $\boldsymbol{x}_{k+1,n}$ are available when recording a data set
Thus, with measurements that reflect the behavior of the autonomous system $n$, the vector $\boldsymbol{w}_{k,n}$ represents the regressors for the least squares method and the vector $\boldsymbol{y}_{k+1,n}$ comprises the values to be predicted by the searched model $\boldsymbol{K}_n$:
\begin{equation}
	\begin{split}
	\boldsymbol{w}_{k,n}\ \ \; &= \begin{bmatrix} \idk & \ \ \ \iqk & \ \ \ \sink & \ \ \; \cosk & \ \ \; 1 \end{bmatrix}^\mathrm{T}, \\
	\boldsymbol{y}_{k+1,n}    &= \begin{bmatrix} \idkk & \iqkk \end{bmatrix}^\mathrm{T}.
	\end{split}
	\label{LS_basic_obervations}
	\end{equation}
	
Afterwards, data matrices $\boldsymbol{X}_{k,n}$ and $\boldsymbol{X}_{k+1,n}$ can be built with $j$ corresponding pairs for each autonomous system $n$:
\begin{equation}
	\begin{split}
	\boldsymbol{W}_{k,n} \ \ \; &= \begin{bmatrix} \boldsymbol{w}_{k,n,1} & \ \ \ \boldsymbol{w}_{k,n,2} & \ \ \ \ldots & \boldsymbol{w}_{k,n,j}\ \ \; \, \end{bmatrix},\\
	\boldsymbol{Y}_{k+1,n}      &= \begin{bmatrix} \boldsymbol{y}_{k+1,n,1} &\ \boldsymbol{y}_{k+1,n,2} & \, \ldots \ \ \ \boldsymbol{y}_{k+1,n,j} \end{bmatrix}.
	\end{split}
	\end{equation}
	
Assuming a sufficiently large set of independent measurements, this leads to an overdetermined system of equations from which the matrix $\boldsymbol{K}_n$ is calculated, with $(\cdot)^+$ denoting the pseudo inverse of a matrix:
\begin{equation}
	\boldsymbol{K}_n = \boldsymbol{Y}_{k+1,n}\boldsymbol{W}_{k,n}^\mathrm{T}\ \bigl(\boldsymbol{W}_{k,n}\boldsymbol{W}_{k,n}^\mathrm{T} \bigr)^+.
	\end{equation}
Using pairs where $\boldsymbol{w}_{k,n}$ is within a defined neighborhood of an operating point results in a prediction model $\boldsymbol{K}_n$ that takes parasitic effects like flux harmonics, inverter nonlinearity or measurement offsets at this operating point implicitly into account.
To use these models in an FCS-MPC, the prediction models would have to be calculated for different operating points and then stored in the controller.

One possible approach to avoid the calculation of different models while still considering the parameter variants of the system, is to extend the vector $\boldsymbol{x}$ by further observations or regressors.
Therefore, it should be pointed out that the regressor configuration of (\ref{LS_basic_obervations}) is only an example of one possible LS-setup. 
However, finding suitable further regressors in the LS framework is not a straightforward way and, therefore, a comprehensive feature engineering should carried out during the pre-processing.

Among other, the SINDy Toolbox can be used to find and analyze additional regressors from a library of possible combinations and functions of the measured quantities $\id$, $\iq$ and $\eel$ \cite{Brunton3932}.\\

\noindent \textbf{Approach c) Machine-learning-based models}\\
The behavior of the plant can also be extracted from data by (deep) machine learning (ML) methods.

For example, supervised learning of artificial neural networks (ANN) can be used for mapping the observations at $k\mathrm{T}_\mathrm{s}$ to the ones at $\left(k+1\right)\mathrm{T}_\mathrm{s}$. 
Later on, these networks are implemented online and used for the prediction of the system states. 
The number of units in the input and output layer is defined by the number of supplied observations. 
The number of hidden layers, the number of neurons per layer, the activation functions and the overall network topology (e.g. feedforward, convolutional, recurrent, ...) are so-called hyperparameters which are the higher level degrees of freedom. 

For a basic ANN the same observations $\boldsymbol{x}_{k,n}$ as for the LS \eqref{LS_basic_obervations} can be used as input to the network. 
However, the constant observation can be omitted.
For the output layer, the two predictions $\idkk$ and $\iqkk$ are sufficient as targets.
Similar to LS, a feature engineering pre-processing may also increase the prediction accuracy.

But also other methods from the domain of machine learning, like Gradient-Boosting-Machines (GBM) or decision trees may be promising approaches.

Especially with ML methods, it is simple to include the information about the vector which was used in the interval before ($n_{k-1}$) as an input.
This might be helpful to consider more detailed effects like the inverter-deadtime or the interlocking time, as they appear when switching between elementary vectors.

\section{Data set}
\label{sec:data set}

For the comparison of the different modeling approaches, a data set including measurements at different operating points is recorded. 
This data set consists of approx. 40 million samples from a defined operating range of the drive.

A sample in the data set (each row) consists of the measured dq-currents at two consecutive time points (e.g. $k$ and $k+1$), the angle at the earlier of the two time points, and the information about the elementary vector selected in the controller cycle between them ($n_k$) as well as the vector selected in the cycle before ($n_{k-1}$). 
An overview of the included variables is given in Tab. \ref{tab:Variables_contained}.
However, the successive rows or samples in the set do not constitute a time series.

\begin{table}[!ht]
\centering
\caption{Variables contained in the data set}
\label{tab:Variables_contained}
\begin{tabular}{c|c|c|c}
\multirow{1}{*}{Variable} & \multirow{1}{*}{Description} & \multirow{1}{*}{Data type} & \multirow{1}{*}{Classification}\\ \hline \hline
$i_{\mathrm{d},k}$        & measured d-current at $k$ & single & \multirow{5}{*}{inputs} \\
$i_{\mathrm{q},k}$        & measured q-current at $k$ & single &                        \\
$\varepsilon_{k}$         & measured rotational angle at $k$ & single &                        \\
$n_k$                     & element. vector applied at $k$ & integer &                        \\
$n_{k-1}$                 & element. vector applied at $k\!-\!1$ & integer &                        \\ \hline
$i_{\mathrm{d},k+1}$      & measured d-current at $k\!+\!1$ & single & \multirow{2}{*}{targets} \\
$i_{\mathrm{q},k+1}$      & measured q-current at $k\!+\!1$ & single &  \\ \hline
\end{tabular}
\end{table}	

As a result of the measurements at $k$ and $k+1$, the real behavior of the currents for a given vector is known.
This knowledge can now be used to derive models.

\begin{figure}[ht]
		\centering
		\includegraphics[width=8.0cm]{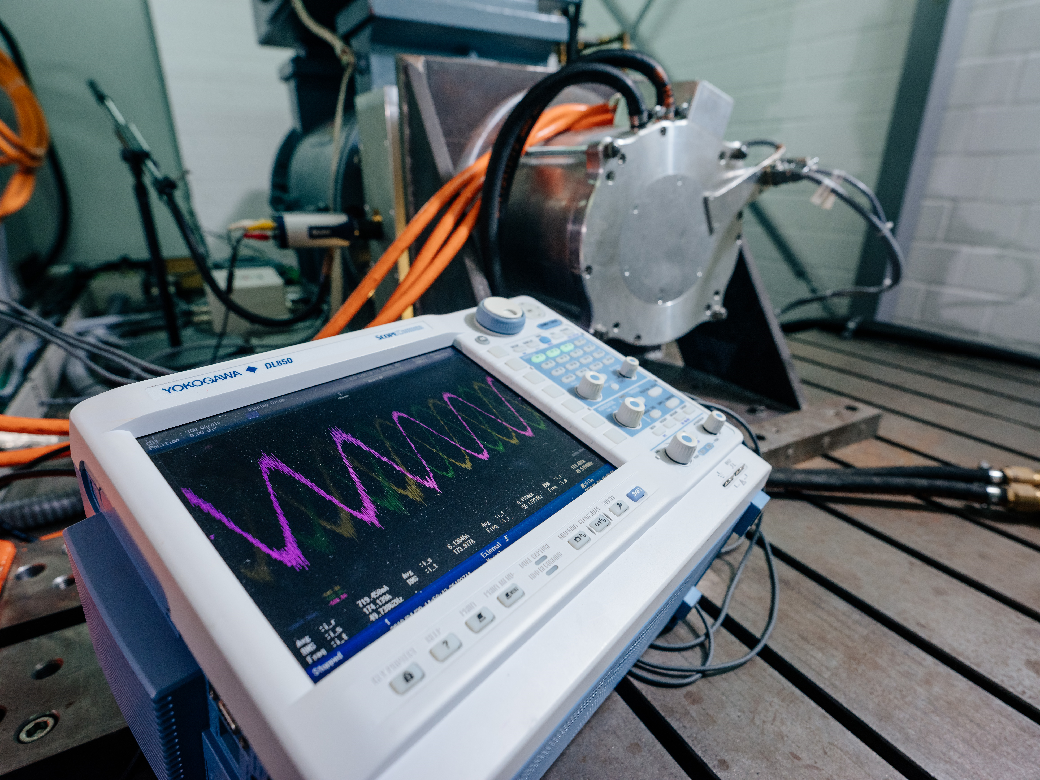}
		\caption{Test bench with the used PMSM in the background}
		\label{fig:test_bench}
	\end{figure}

The drive system under test consists of an interior magnet permanent magnet synchronous motor (IPMSM) of \SI{57}{\kW} and a 2-level IGBT inverter. 
The most important test bench parameters are summarized in Tab. \ref{tab:param}.
Fig. \ref{fig:test_bench} shows the test bench with the transient recorder in the front and the used motor in the background.

\begin{table}[ht]
				\caption{Test bench parameters}
			  \label{tab:param}
				\centering
				\begin{tabular}{l|c|c}
				\hline
				\textbf{DC Power supply} & \multicolumn{2}{c}{Gustav Klein} \\
				DC output                   & \multicolumn{2}{c}{galvanically isolated} \\
				Max. apparent power					& $S_\mathrm{max}$ & 200\,kW \\
				Max. DC current             & $I_\mathrm{DC,max}$ & 500\,A \\
				Variable DC output voltage  & $U_\mathrm{DC}$    & 6-600\,V \\ \hline \hline
				\textbf{IPMSM} & \multicolumn{2}{c}{Brusa HSM16.17.12-C01} \\
				Stator resistance         & $R_\mathrm{s}$    & 18\,m$\Omega$ 						\\ 
				Inductance in d-direction & $L_\mathrm{d}$    & \SI{370}{\micro\henry}    \\ 
				Inductance in q-direction & $L_\mathrm{q}$    & \SI{1200}{\micro\henry}   \\
				Permanent magnet flux     & $\Psip$           & 66\,mV\,s                 \\
				Pole pair number          & $p$               & 3                         \\
				Rated machanical power    & $P_\mathrm{me}$   & 57\,kW                    \\
				Rated torque              & $M$               & 130\,N\,m                 \\ 
				Max. stator current in dq-system & $|\idq|_\mathrm{max}$   & 240\,A                    \\ \hline \hline
				\textbf{Inverter} & \multicolumn{2}{c}{3$\times$SKiiP 1242GB120-4D} \\
				Typology                  & \multicolumn{2}{c}{voltage source inverter} \\
				        									& \multicolumn{2}{c}{2-level, IGBT} \\
				Max. phase current        & $I_\mathrm{C,max}$ & 1200\,A \\ \hline \hline
				\textbf{Controller hardware} & \multicolumn{2}{c}{dSPACE} \\
				Processor board              & \multicolumn{2}{l}{DS1006MC, 4 cores, 2.8\,GHz} \\
				FPGA board              & \multicolumn{2}{l}{DS5203, Xilinx Virtex-5} \\
				ADC board               & \multicolumn{2}{l}{DS2004, 16 channel, 16 bit} \\
				PWM board               & \multicolumn{2}{l}{DS5101} \\
				CAN board               & \multicolumn{2}{l}{DS4302} \\
				Digital I/O board       & \multicolumn{2}{l}{DS4003} \\
				Incremental encoder board  & \multicolumn{2}{l}{DS3002} \\ \hline \hline
				\textbf{Measurement devices} & \multicolumn{2}{c}{}  \\
				Transient recorder & \multicolumn{2}{c}{Yokogawa\ \ \ DL850} \\
				Power analyzer     & \multicolumn{2}{c}{Yokogawa WT3000} \\
				                   %& \multicolumn{2}{c}{\ 2$\times$Hioki\ \ PW6001} \\
				Current probes                         & \multicolumn{2}{l}{4$\times$Danfysik, 700\,A, 100\,kHz} \\
				(all zero-flux transducers)					   & \multicolumn{2}{l}{3$\times$Yokogawa, 500\,A, 2\,MHz}  \\
																						%	 & \multicolumn{2}{l}{12$\times$Hioki,\ \ \ \ \ \, 500\,A, 100\,kHz} \\
				                                    %  & \multicolumn{2}{l}{11$\times$Yokogawa, 500\,A,\ \ \ \ 2\,MHz}  \\	
																						%	 & \multicolumn{2}{l}{\ \,4$\times$Iwatsu,\ \ \ \ \ \; 30\,A, 100\,MHz} \\
				Torque sensors 												 & \multicolumn{2}{l}{\ \ \ \ \, HBM, T10FS, 2\,kN\,m} \\ \hline 
																							 %& \multicolumn{2}{l}{\ \ \ \ \, HBM, T10F, \; 1\,kN\,m} \\ 
																							 %& \multicolumn{2}{l}{\ \ \ \ \, HBM, T12, \, \: 1\,kN\,m} \\ \hline 
				\end{tabular}
		\end{table}	
		
The rotational speed of the motor and the DC link voltage for all samples were constant at $n_\mathrm{me}=1000$\,min$^{-1}$ and $U_\mathrm{DC}=300$\,V.
Hence, these variables are not part of the data set. 
In the future, an extended data set for varying rotational motor speeds may be added and then the rotational speed would be added to the input space. 
Similar, the motor temperature was nearly constant during all measurements and, therefore, does not need to be considered in the given data set.
The parameters that are specific for this data set are summarized in Tab. \ref{tab:drive_train_param}.

%\noindent \textbf{Please note:}\\
%For sake of simplicity, $\UDC$ is considered as ideally constant in this contribution.
%Moreover, the rotational speed $n_\mathrm{me}$ is kept constant.
%It is planed to extend the data set to variations of this two parameters in the future.
%However, the presented data driven modeling approaches can be directly extended to consider a varying DC-link voltage and rotational speed.

\begin{table}[ht]
				\caption{Drive train parameters}
			  \label{tab:drive_train_param}
				\centering
				\begin{tabular}{l|c|c}
				\hline
				Mechanical speed										& $n_\mathrm{me}$  & 1000\,min$^{-1}$ \\
				DC-link voltage           					& $U_\mathrm{DC}$  & 300\,V \\
				Stator temperature         					& $\vartheta_\mathrm{s}$  & \SI{55}{\celsius} \\
				FCS-MPC: Controller cycle time   	  & $\mathrm{T_s}$ 	 & \SI{50}{\us} \\
				FCS-MPC: Max. switching frequency   & $f_\mathrm{sw}$	 & 10\,kHz\\ 
			  FCS-MPC: Prediction horizon    		  & $n_\mathrm{p}$   & 1\\ \hline
				\end{tabular}
		\end{table}		
		
\begin{figure}[ht]
		\centering
		\includegraphics[]{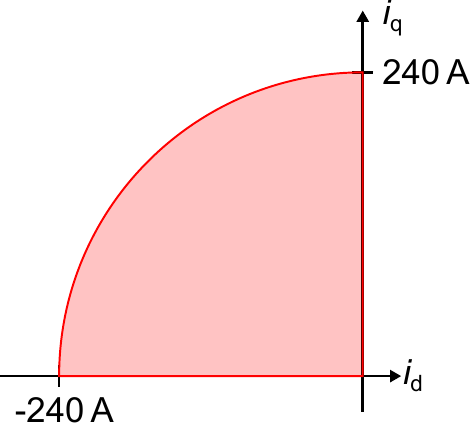}
		\caption{Defined current operating range in the dq-plane}
		\label{fig:operating_range}
	\end{figure}

Besides the constant variables, the operating range is defined by a variation of the $\idq$ currents within the shown quadrant of the dq-plane (Fig. \ref{fig:operating_range}). For this motor, the maximum allowed length of the $\idq=\begin{bsmallmatrix} \id & \iq \end{bsmallmatrix}^\top$ current vector is \SI{240}{\ampere}. The value of the rotor angle $\eel$ ranges from -$\pi$ to $\pi$.
%\SI[explicit-sign = -]{180}{\degree} to \SI{180}{\degree}.\\

In order to record measurements from the entire operating range, a sequence of 478 different $\idq$ set points was commanded to the FCS-MPC.
At each resulting $\idq$ operating point, a few seconds of measurement data were then recorded.
To capture the behavior for an extensive combination of $\vn$ and $\eel$ at a given $\idq$ operating point, some of the elementary vectors selected by the MPC were replaced by randomly selected vectors. 

\subsection{Balancing of the data set}		
		
If a model is trained for the whole operating range, a homogeneous distribution of the measurements in the data set over the operating range of $\id$, $\iq$ and $\eel$ is important.
This ensures that the model is not biased towards regions in the operating range where the sample concentration is higher resulting in a reduced accuracy in regions with a minor sample concentration \cite{MacNamee.2002}. This generally applies to all kinds of models that are learned or built on the basis of data and are intended to cover the entire operating range.

One method to obtain a balanced data set is described in the following.
First, the operating range can be divided into classes by a grid. 
The grid step size and the operating range that are used for this example are summarized in Tab. \ref{tab:Step_sizes}.
The number of samples per class and, thus, the balance of the data set can then be analyzed.

\begin{table}[!ht]
\centering
\caption{Grid step sizes and operating range}
\label{tab:Step_sizes}
\begin{tabular}{c|c|ccc|c}
\multicolumn{2}{c|}{Dimension}           & \multicolumn{3}{c|}{Operating range} & Grid step size \\ \hline \hline
d-current                       & $\id$  & -240\,A & to & 0\,A        & 10\,A \\ 
q-current                       & $\iq$  &    0\,A & to & -240\,A     & 10\,A \\
rotation angle                  & $\eel$ & -$\pi$   & to & $\pi$      & $\pi/18$ \\
\end{tabular}
\end{table}			

\begin{figure}[ht]
		\centering
		\includegraphics[]{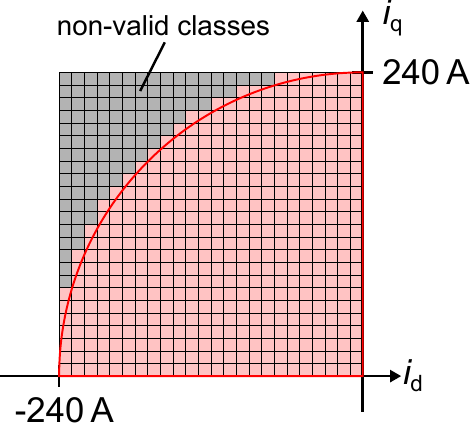}
		\caption{Classes within the defined current operating range in the dq-plane, of which 471 are valid}
		\label{fig:operating_range_valid_classes}
	\end{figure}

\begin{figure}[ht]
		\centering
		\includegraphics[]{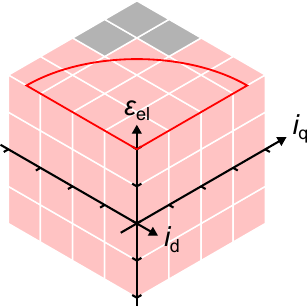}
		\caption{Valid classes within the defined current and angle operating range (simplified representation)}
		\label{fig:operating_range_valid_classes_cube}
	\end{figure}

For the dq-plane, the grid is shown in Fig. \ref{fig:operating_range_valid_classes}.
Valid classes are all classes that are fully or partially within the specified current operating range, they are shaded red.
This pattern continues for the range of the rotor angle as shown in Fig. \ref{fig:operating_range_valid_classes_cube}.

Furthermore, the data set is divided into subsets where each subset comprises the samples of one elementary vector $\vn$.
This provides the opportunity to extract a model for each autonomous system $n$, as already described in Sec. \ref{sec:Prediction_models} for the LS approach.

\begin{figure}[ht]
		\centering
		\includegraphics[]{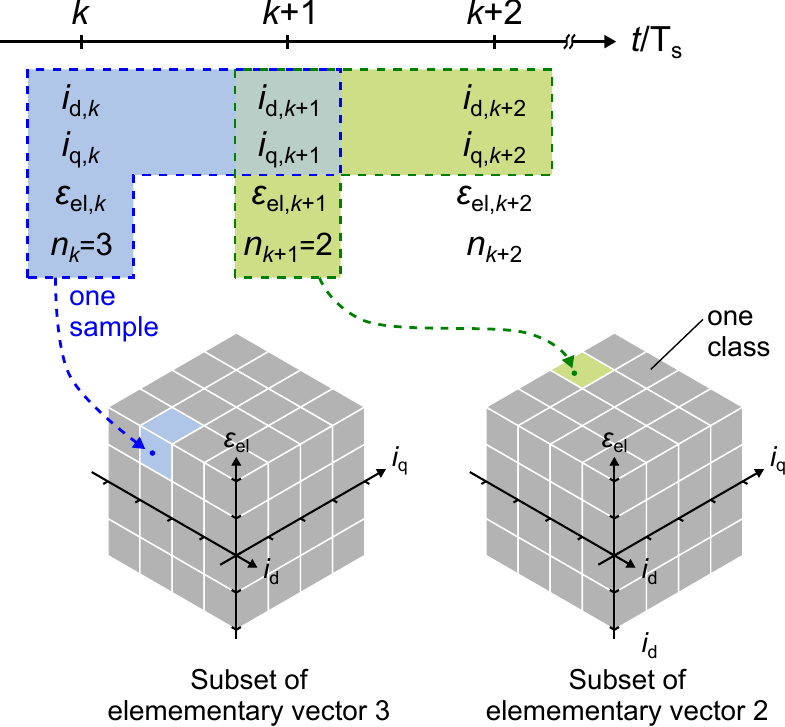}
		\caption{Assignment of samples to subsets and classes}
		\label{fig:sequence}
	\end{figure}

Fig. \ref{fig:sequence} shows the assignment of recorded measurements to samples, subsets and classes within the subsets.
A sample consists of the dq-currents of two consecutive time points, the angle at the earlier of the two time points, and the information about the chosen elementary vector.
The assignment of a sample to a subset is done by means of the elementary vector which is applied between the two time points.
The class assignment of a sample is determined according to the currents and the angle at the earlier of the two points in time.
For the shown blue sample, the class to which this sample belongs is determined by the vector $\begin{bsmallmatrix} \idk & \iqk & \eelk \end{bsmallmatrix}^\top$.
		
\begin{figure}[ht]
		\centering
		\includegraphics[]{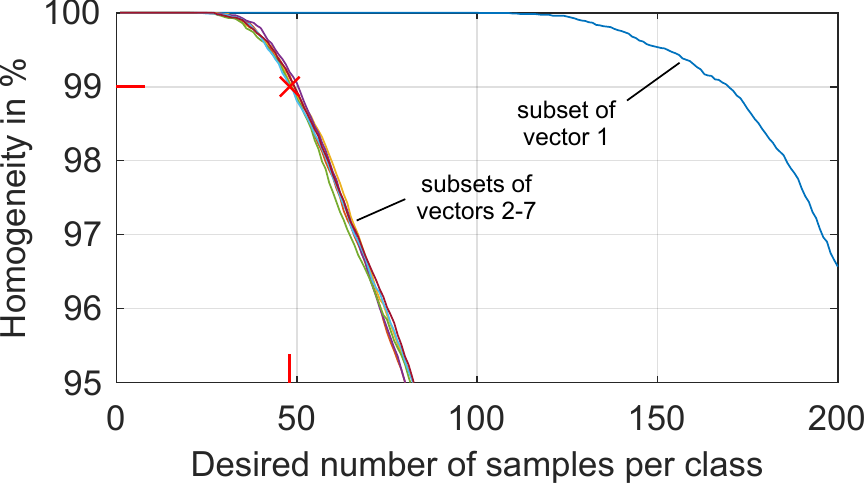}
		\caption{Homogeneity of the data set for a desired number of samples per class for each subset with a grid step size of \SI{10}{\ampere} for $\id$, $\iq$ and $\pi/18$ for $\eel$, resulting in a number of 16956 ($=471\cdot$36) valid classes per subset}
		\label{fig:Abdeckung}
	\end{figure}

After the assignment of all available samples, the distribution of samples regarding the valid classes can be analyzed. 
This is done for each subset.
Fig. \ref{fig:Abdeckung} shows the homogeneity for each subset in dependence of the number of samples per class.
As an example, if the desired number of samples per class is set to 48 samples, \SI{99}{\percent} of the valid classes meet this requirement because they contain more or exactly this desired number of samples (red marker).

Thus, limiting the number of samples to 48 in each class, leads to nearly homogenous (\SI{99}{\percent}) data set which can be used for the training of an LS or an ANN. 
If a class contains more than 48 samples, the remaining samples are transferred to an non-homogeneous data set that can be used to test the learned models on samples that were not utilized during learning or built-up.

Since elementary vector $\boldsymbol{v}_1$ was selected relatively often by the controller when recording the data set, considerably more samples are available per class.
The frequent selection of this so-called zero voltage vector results from the chosen value of DC link voltage in combination with the operating range, especially with the rotational speed.
Even with a higher number of desired samples, there is still a high degree of homogeneity for this subset as can be seen from the blue curve.

\noindent \textbf{Please note:}\\
For sake of simplicity, $\UDC$ is considered as ideally constant in this contribution.
Moreover, the rotational speed $n_\mathrm{me}$ and the motor temperature are kept constant, too.
It is planed to extend the data set to variations of this three variables in the future.
However, the presented data-driven modeling ideas can be directly extended to consider these varying operation conditions by extending the input space with this additional features. \newline

\noindent \textbf{Link to the uploaded data set:}\\
The data set is published on Kaggle, an online community of data scientists: \url{https://www.kaggle.com/hankelea/system-identification-of-an-electric-motor}

\bibliography{refs}

\end{document}